\documentclass[aps,prd,nofootinbib,twocolumn,showpacs,superscriptaddress,groupedaddress,preprintnumbers,floatfix]{revtex4-1}  % for review and submission
\usepackage{graphicx}  % needed for figures
\usepackage{dcolumn}   % needed for some tables
\usepackage{bm}        % for math
\usepackage{amssymb}   % for math
\usepackage{subfigure}
\usepackage{multirow}
\usepackage{placeins}
\usepackage{epstopdf}
\usepackage{amsfonts}    
\usepackage{amsmath}
\usepackage{color}

\begin{document}

\preprint{
\vbox{
\hbox{MIT-CTP/4811}
\hbox{INT-PUB-16-017}
}}

\widetext

\title{Gluonic Transversity from Lattice QCD}

\author{W.~Detmold}\affiliation{Center for Theoretical Physics, Massachusetts Institute of Technology, Cambridge, MA 02139, U.S.A.}
\author{P.~E.~Shanahan}\affiliation{Center for Theoretical Physics, Massachusetts Institute of Technology, Cambridge, MA 02139, U.S.A.}

\begin{abstract}
We present an exploratory study of the gluonic structure of the $\phi$ meson using lattice QCD (LQCD). This includes the first investigation of gluonic transversity via the leading moment of the twist-two double-helicity-flip gluonic structure function $\Delta(x,Q^2)$. This structure function only exists for targets of spin $J\ge1$ and does not mix with quark distributions at leading twist, thereby providing a particularly clean probe of gluonic degrees of freedom. We also explore the gluonic analogue of the Soffer bound which relates the helicity flip and non-flip gluonic distributions, finding it to be saturated at the level of 80\%. This work sets the stage for more complex LQCD studies of gluonic structure in the nucleon and in light nuclei where $\Delta(x,Q^2)$ is an `exotic glue' observable probing gluons in a nucleus not associated with individual nucleons.
\end{abstract}


\maketitle

\section{Introduction}
\label{sec:Introduction}

Quantitatively describing the structure of hadrons, especially the nucleon, in terms of the quark and gluon constituents encoded in Quantum Chromodynamics (QCD) is a defining challenge for hadronic physics. The ultimate goal is to map the complete spatial, momentum, spin, flavour, and gluon structure of hadrons. Such a map is not only the key to interpreting our observations of Nature in terms of the currently-accepted fundamental theory, but is essential to inform searches for physics beyond the Standard Model at both the high-energy and intensity frontiers. 
While many observables related to quark distributions in hadrons have been measured and studied~\cite{DeRoeck:2011na,Agashe:2014kda}, gluon distributions have received less attention, in part because of the experimental challenges inherent in measurements of these quantities.
As a primary mission of the proposed Electron-Ion Collider~\cite{Accardi:2012qut,Kalantarians:2014eda} is to study glue in the proton and in nuclei, significant experimental progress may be expected on this front over the next decade. There is also potential for experiments to study gluon distributions at Jefferson Lab~\cite{JLAB.LOI} and at the LHC~\cite{Baltz:2007kq}.

In this work, we study the gluon structure of the spin-1 $\phi(\overline{s}s)$ meson through a calculation of the first moments of its spin-independent and transversity distributions. This constitutes the first lattice QCD calculation of the leading-twist, double-helicity-flip transversity structure function, named $\Delta(x,Q^2)$~\cite{Jaffe:1989xy}, in any hadron. 
This quantity is particularly interesting since, unlike the unpolarised and helicity gluon distributions, the double-helicity-flip density is a clean measure of gluonic degrees of freedom as it only mixes with quark distributions at higher twist.
The only existing information on $\Delta(x,Q^2)$ comes from a rudimentary bag model calculation of its first moment in the spin-$\frac{3}{2}$ $\Delta$ baryon~\cite{Sather:1990bq}, and a related model of its $x$-dependence in the deuteron~\cite{Nzar:1992ax}. 
We also study the gluonic analogue of the Soffer bound for transversity for the first time, showing that the first moment of this bound in a $\phi$ meson (at the unphysical light quark masses used in this work and subject to caveats regarding renormalisation and the continuum limit) is saturated to approximately the same extent as the first moment of the quark Soffer bound for the nucleon as determined in a previous lattice simulation~\cite{Gockeler:2005cj,Diehl:2005ev}. 

This work demonstrates that complex aspects of gluonic structure are accessible to lattice QCD calculations (previously the unpolarised gluonic structure of the pion and nucleon have been investigated~\cite{Meyer:2007tm,Horsley:2012pz,Alexandrou:2013tfa}).
It also lays the groundwork for extensions in several phenomenologically-interesting directions. While the nucleon has no gluon transversity distribution at twist-2, its helicity-flip off-forward parton distributions are non-vanishing. 
These quantities, which can be calculated on the lattice using methods similar to those discussed in this work, can be probed through distinct angular dependence of the cross-section in deeply-virtual Compton scattering (DVCS)~\cite{Hoodbhoy:1998vm}.
In nuclei with spin $J\ge1$, it has been recognised since 1989~\cite{Jaffe:1989xy} that the structure function $\Delta (x,Q^2)$ is sensitive to exotic glue---the contributions from gluons not associated with individual nucleons in a nucleus---as neither nucleons nor pions (nor anything with spin less than one) can transfer two units of helicity to the nuclear target. This structure function can be measured in deep inelastic scattering (DIS) on spin $J\ge 1$ targets, as has been proposed for nitrogen targets in a recent letter of intent to Jefferson Lab~\cite{JLAB.LOI}.

\section{Definition of $\Delta (x,Q^2)$}

The observable $\Delta (x,Q^2)$ was introduced in Ref.~\cite{Jaffe:1989xy} as a new leading-twist structure function which can be measured in deep inelastic scattering from polarized spin $\ge 1$ targets. We follow that reference in defining and outlining the construction of $\Delta (x,Q^2)$ below.

The hadronic part of the differential cross section for inelastic lepton scattering from a polarized spin-one target can be expressed as
\begin{equation}
W_{\mu\nu}(p,q,E',E) = \frac{1}{4\pi} \int d^4x\, e^{iq\cdot x} \langle p',E' | [j_\mu(x),j_\nu(0)]|p,E\rangle,
\end{equation}
where $E_\mu^{(\prime)}$ is a polarization vector describing the spin orientation of the target, with $p^{(\prime)}\cdot E^{(\prime)}=0$ and $E^{(\prime)2}=-1$. The target four-momentum is denoted $p^\mu$ while $q^\mu$ denotes the transferred four-momentum to the target.
The dependence of this expression on the polarizations $E$ and $E'$ can be factored out to define a target-polarization--independent tensor $W_{\mu\nu,\alpha\beta}$:
\begin{equation}
W_{\mu\nu}(p,q,E,E')=E'^{*\alpha} E^\beta W_{\mu\nu,\alpha\beta}(p,q).
\end{equation}

The tensor $ W_{\mu\nu,\alpha\beta}$ can be related to helicity projection operators $P(hH,h'H')_{\mu\nu,\alpha\beta}=\epsilon^*_{\mu} (h') E^*_{\alpha} (H')E_{\beta}(H)\epsilon_{\nu} (h)$, where $\epsilon_{\mu} (h)$ are photon polarization vectors and the helicity components of the photon and target are denoted $h$ and $H$:
\begin{equation}\label{eq:Wmunualbet}
W_{\mu\nu,\alpha\beta}(p,q)=\sum_{hH,h'H'}P(hH,h'H')_{\mu\nu,\alpha\beta}A_{hH,h'H'}(p,q).
\end{equation}
Here $A_{hH,h'H'}$ represents the imaginary part of the corresponding forward Compton helicity amplitude. Writing the double-helicity-flip component $A_{+-,-+}=A_{-+,+-}$ (where the equality follows by parity invariance) as $\Delta(x,Q^2)$, the double-helicity-flip part of Eq.~\eqref{eq:Wmunualbet} becomes
\begin{equation}
W_{\mu\nu,\alpha\beta}^{\Delta=2}=\left[P(+-,-+)_{\mu\nu,\alpha\beta}+P(-+,+-)_{\mu\nu,\alpha\beta}\right]\Delta(x,Q^2).
\end{equation}

\begin{figure}
	\includegraphics[width=0.35\textwidth]{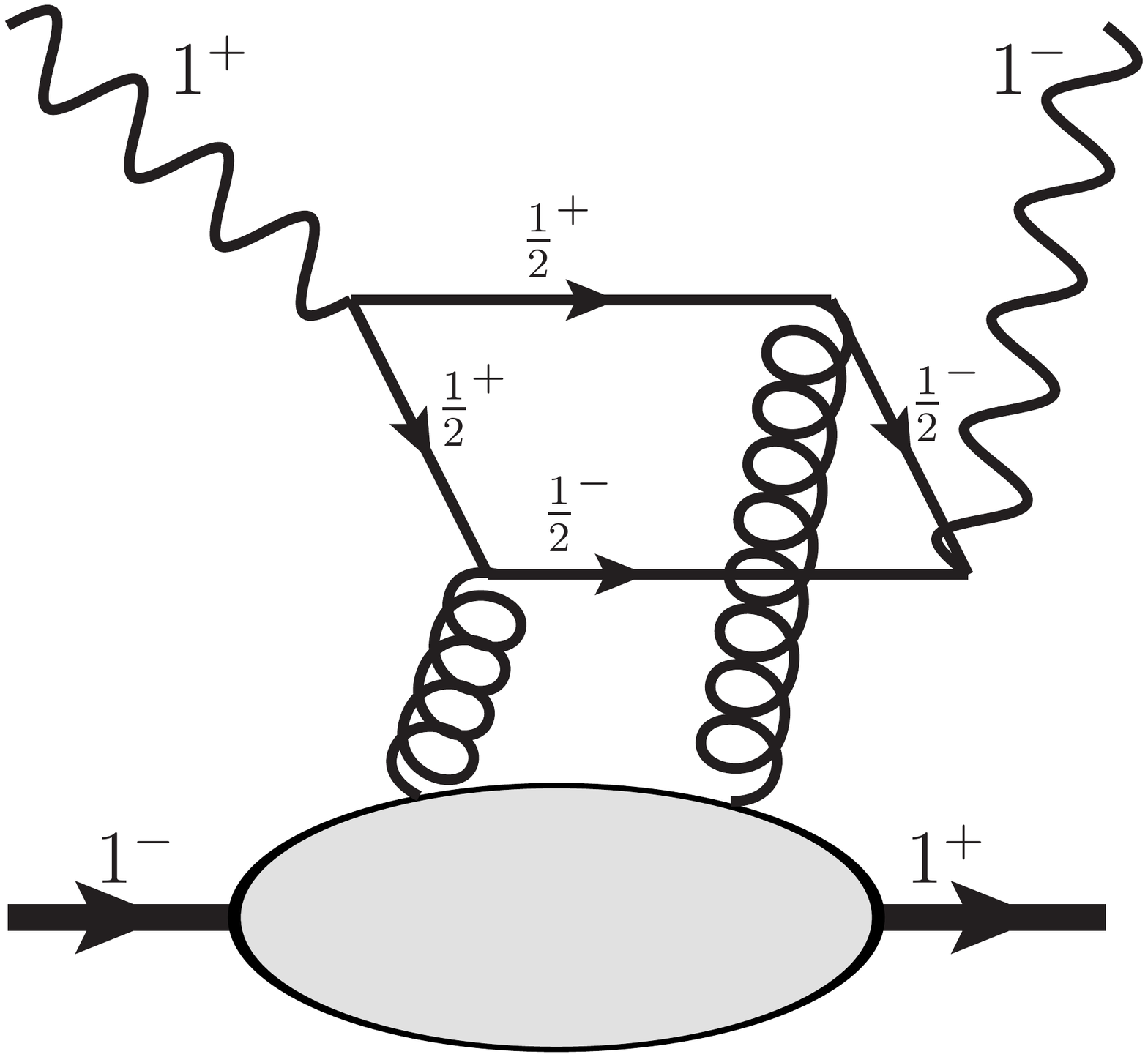}
	\caption{\label{fig:DIS}Illustration of one of the leading contributions in DIS sensitive to $\Delta(x,Q^2)$. The wavy, curly and thin lines denote photons, gluons and quarks, while the thick line represents a spin-1 hadron.}
\end{figure}

Finally, expanding the helicity projection operators explicitly in terms of the photon and target polarization vectors, the double-helicity-flip term in $W_{\mu\nu}(p,q)$ can be expressed as~\cite{Jaffe:1989xy}:
\begin{widetext}
\begin{align}\nonumber
W_{\mu\nu}^{\Delta=2}=\frac{1}{2}\left\{\vphantom{\frac{q\cdot E'^*}{\kappa\nu}}\right.&\left[\left(E'^*_\mu-\frac{q\cdot E'^*}{\kappa\nu}\left(p_\mu - \frac{M^2}{\nu}q_\mu\right)\right)\left(E_\nu-\frac{q\cdot E}{\kappa\nu}\left(p_\nu - \frac{M^2}{\nu}q_\nu\right)\right)+(\mu\leftrightarrow \nu)\right]\\
&-\left.\left[g_{\mu\nu}-\frac{q_\mu q_\nu}{q^2}+\frac{q^2}{\kappa\nu^2}\left(p_\mu-\frac{\nu}{q^2}q_\mu\right)\left(p_\nu-\frac{\nu}{q^2}q_\nu\right)\right]\left[E'^*\cdot E + \frac{M^2}{\kappa \nu^2}q\cdot E'^* \,q\cdot E\right]\right\}\Delta(x,Q^2).
\end{align}
\end{widetext}
This expression will vanish if $E'=E$ or if averaged over spin.
The leading contribution in DIS sensitive to $\Delta(x,Q^2)$ is illustrated in Fig.~\ref{fig:DIS}.

To relate $\Delta(x,Q^2)$ to matrix elements of operators in the operator product expansion, we consider the time-ordered product of two vector currents, 
\begin{equation}
\mathcal{T}_{\mu\nu}(q)\equiv i \int d^4x\, e^{iq\cdot x} T(j_\mu(x)j_\nu(0)),
\end{equation}
and perform an operator product expansion (OPE) near the lightcone.
At leading twist (twist-2), the only contribution which does not vanish when contracted with $P(\pm\mp,\mp\pm)_{\mu\nu,\alpha\beta}$, and can therefore contribute to the double-helicity-flip Compton amplitude, arises from a tower of gluonic operators:
\begin{equation}\label{eq:Tbit}
\frac{1}{2}\mathcal{T}_{\mu\nu}^{\Delta=2}(q) = \sum_{n=2,4,\ldots} \frac{2^n q^{\mu_1}\ldots q^{\mu_n}}{(Q^2)^n}C_n(Q^2,\mu^2)\mathcal{O}_{\mu\nu\mu_1\ldots\mu_n}(\mu^2),
\end{equation}
where $\mu$ is the factorization and renormalisation scale,
\begin{equation}\label{eq:operator}
\mathcal{O}_{\mu\nu\mu_1 \ldots \mu_{n}} (\mu^2)= S\left[G_{\mu\mu_1} \overleftrightarrow{D}_{\mu_3}\ldots \overleftrightarrow{D}_{\mu_{n}}G_{\nu \mu_2}\right],
\end{equation}
and throughout this paper `$S[\,]$' denotes symmetrisation in the indices $\mu_1, \ldots, \mu_n$ and trace-subtraction in all indices.
$G_{\mu\nu}$ is the gluon field strength tensor and $D_\mu$ denotes the gauge covariant derivative.
The Wilson coefficients in the OPE are\footnote{This expression agrees with those in Refs.~\cite{Hoodbhoy:1998vm,Belitsky:2000jk}, but differs by a sign from that in Ref.~\cite{Jaffe:1989xy}.}
\begin{equation}
C_n(Q^2,\mu^2)= -\frac{\alpha_s(Q^2)}{2\pi}\textrm{Tr} \mathcal{Q}^2 \frac{2}{n+2},
\end{equation}
where $\mathcal{Q}$ denotes the quark charge matrix and at leading order there is no dependence on the factorization scale. 

In a spin-one target with polarization $E$ and $E'$, the forward matrix element of the operator $\mathcal{O}$ is
\begin{align} \nonumber
\langle p E' | \mathcal{O}_{\mu\nu\mu_1\ldots\mu_n}|p E\rangle \\\nonumber
= (-2i )^{n-2}\frac{1}{2}
S\left[ \left\{\vphantom{E'} \right. \right. &( p_\mu E'^{*}_{\mu_1} - p_{\mu_1}E'^{*}_\mu)(p_\nu E_{\mu_2}-p_{\mu_2}E_\nu)\\ \label{eq:expe}
  & + \left. \left.\vphantom{E'} (\mu \leftrightarrow \nu) \right\}p_{\mu_3} \ldots p_{\mu_n} \right]A_n(\mu^2),
\end{align}
where `$S$' is as above\footnote{This definition of $A_n$ differs from that in Ref.~\cite{Jaffe:1989xy} by a factor of two, chosen for convenience for the discussion of the Soffer bound in this work.}.

The reduced matrix elements $A_n$, for even $n$, can be related to moments of the structure function $\Delta(x,Q^2)$. Writing the subtracted dispersion relation for the double-helicity-flip part of the matrix element of $\mathcal{T}_{\mu\nu}$ (Eq.~\eqref{eq:Tbit}) and using the optical theorem to relate the imaginary part of the matrix element of $\mathcal{T}_{\mu\nu}$ to $W_{\mu\nu}$ (and hence to $\Delta(x,Q^2)$) gives the identification
\begin{equation}\label{eq:Mn}
M_n(Q^2) = C_n(Q^2,Q^2)\frac{A_n(Q^2)}{2}, \hspace{4mm}n=2,4,6\ldots,
\end{equation}
where $A_n$ is renormalized at the scale $\mu^2=Q^2$, and
\begin{equation}
M_n(Q^2) = \int_0^1 dx x^{n-1} \Delta(x,Q^2)
\end{equation}
are the Mellin moments of $\Delta(x,Q^2)$.

The structure function $\Delta(x,Q^2)$ also has a parton model interpretation. For a target in the infinite momentum frame polarized in the $\hat{x}$ direction perpendicular to its momentum (defined to be in the $\hat z$ direction),
\begin{equation}
\Delta(x,Q^2) = -\frac{\alpha_s(Q^2)}{2\pi}\textrm{Tr}\mathcal{Q}^2\,x^2 \int_x^1 \frac{dy}{y^3}\,\delta G(y,Q^2),
\end{equation}
where $\delta G$ is again renormalized at the scale $\mu^2=Q^2$, and
\begin{equation}\label{eq:deltaG}
\delta G(x,\mu^2) = g_{\hat{x}}(x,\mu^2)-g_{\hat{y}}(x,\mu^2).
\end{equation}
Here $g_{\hat{x},\hat{y}}(x,\mu^2)$ denotes the probability of finding a gluon with longitudinal momentum fraction $x$ linearly polarized in either of the transverse directions, $\hat{x}$ or $\hat{y}$, in the transversely polarized target.

\section{Lattice Calculations}
\label{sec:lattice}

In order to calculate the reduced matrix elements $A_n$  appearing in Eqs.~\eqref{eq:expe} and \eqref{eq:Mn} using lattice QCD, we must calculate the expectation values of local operators of the form of Eq.~\eqref{eq:operator}. Here we describe these lattice calculations, discuss the construction of appropriate Euclidean-space local operators for the $n=2$ case, and summarize the methods used to extract the corresponding reduced matrix element $A_2$. Since this is an exploratory calculation, it is performed at a single set of lattice parameters and a number of systematic issues are left to future work.

\subsection{Lattice Simulation}

\begin{table*}
\begin{tabular}{ccccccccccccccc}\toprule
$L/a$ & $T/a$ & $\beta$ & $am_l$ & $am_s$ & $a$~(fm) & $L$~(fm) & $T$~(fm) & $m_\pi$~(MeV) & $m_K$~(MeV) & $m_\phi$~(MeV) & $m_\pi L$ & $m_\pi T$ & $N_\textrm{cfg}$ & $N_\textrm{src}$ \\ \hline
24 & 64 & 6.1 & -0.2800 & -0.2450 & 0.1167(16) & 2.801(29) & 7.469(77) & 450(5) & 596(6) & 1040(3) & 6.390 & 17.04 & 1042 & $10^5$ \\\toprule
\end{tabular}
\caption{\label{tab:configs}Parameters of the ensemble of gauge-field configurations. The lattices have dimension $L^3\times T$, lattice spacing $a$ and bare quark masses $a m_q$ (in lattice units). A total of $N_\textrm{src}$ light-quark sources were used to perform measurements on $N_\textrm{cfg}$ configurations.}
\end{table*}

Calculations were performed on an ensemble of  isotropic gauge-field configurations with $N_f=2+1$ flavours of dynamical quarks. Specifics of this ensemble are given in Table~\ref{tab:configs}~\cite{PhysRevD.92.114512}. The lattices have dimensions $L^3\times T=24^3\times 64$ with lattice spacing $a=0.1167(16)$~fm. 
The L\"uscher-Weisz gauge action~\cite{LuscherWeisz} was employed with a clover-improved quark action~\cite{Sheikholeslami:1985ij} with one level of stout link smearing ($\rho=0.125$)~\cite{PhysRevD.69.054501}. The clover coefficient was set equal to its tree-level tadpole-improved value.
The light quark masses are such that the pion mass is 450(5)~MeV
and the strange quark mass is such that the resulting mass of the $\phi$ is 1040(3)~MeV.

\subsection{Lattice Operator Construction}
\label{sec:lattop}

In this work we consider the lowest dimension ($n=2$) operator of the tower in Eq.~\eqref{eq:operator}:
\begin{equation}\label{eq:O2}
\mathcal{O}_{\mu\nu\mu_1 \mu_{2}} = S\left[G_{\mu\mu_1}G_{\nu \mu_2}\right].
\end{equation}
The symmetrized and trace-subtracted operator transforms irreducibly as $(2,2)$ under the Lorentz group and does not mix with quark-bilinear operators of the same dimension under renormalization (this operator mixes into higher twist four-quark operators, but the reverse mixing is highly suppressed).
On a hypercubic lattice, the Lorentz group is reduced to the hypercubic group $H(4)$, increasing the possibilities for operator mixing.

Lattice operators with the appropriate continuum behavior that are safe from mixing with lower or same-dimensional operators can be constructed by considering their symmetry properties under $H(4)$. 
The basic building block of such operators is
\begin{equation}\label{eq:OE}
\mathcal{O}^{(E)}_{\mu\nu\mu_1 \mu_{2}} = G^{(E)}_{\mu\mu_1}G^{(E)}_{\nu \mu_2},
\end{equation}
where symmetrisation of indices is not implied.

The transformation properties of quark operators with the symmetries of Eq.~\eqref{eq:OE} under $H(4)$ were described, for the $n=2$ case, in Ref.~\cite{Gockeler:1996mu}. 
We use the same notation as in that work, with the 20 inequivalent irreducible representations of $H(4)$ denoted by $\tau_k^{(d)}$ where $d$ denotes the dimension of the representation and $k$ distinguishes between inequivalent representations of the same dimension. Using the embedding of $H(4)$ into $GL(4)$ to classify the symmetry properties of each irreducible representation, the bases of interest here (i.e., those which have the same symmetry as the operator under consideration) are those in the irreducible subspace corresponding to a $2\times 2$ Young frame. 

The symmetry properties of operators which could possibly mix with $\mathcal{O}^{(E)}_{\mu\nu\mu_1 \mu_{2}}$ are given in Table~\ref{tab:1} in terms of the defining representation labelled as $\tau_1^{(4)}$ and the odd-parity representation labelled as $\tau_4^{(4)}$. %Clearly, the dimension-4 rank-4 operator $\mathcal{O}^{(E)}_{\mu\nu\mu_1 \mu_{2}}$ does not mix with quark operators of the same dimension as these have rank 2.

Table~\ref{tab:irrepAppear} shows the rank at which irreducible representations first appear in each tower of tensor products (decomposed into direct sums)  in Table~\ref{tab:1}. Of the representations that first appear at rank $m=4$ (corresponding to the $n=2$ operator), $\tau_1^{(2)}$, $\tau_2^{(2)}$, and $\tau_2^{(6)}$ also appear in the $GL(4)$-irreducible subspace which has the correct symmetries for Eq.~\eqref{eq:O2}. We therefore choose to consider lattice operators transforming under these three irreducible representations as they cannot mix with any quark or gluon operators of the same or lower dimension. Explicit forms of the ten (2+2+6 from the three representations) basis vectors we consider are given in Appendix~\ref{sec:latticeExplicit}.

To implement the lattice operator, $\mathcal{O}_{\mu\nu\mu_1\mu_2}^{\text{latt.}}$, 
we use the clover definition of the field strength tensor
\begin{equation}
G_{\mu\nu}(x) = \frac{1}{4}\frac{1}{2}\left(P_{\mu\nu}(x)-P^\dagger_{\mu\nu}(x)\right),
\end{equation}
where
\begin{align}\nonumber
P_{\mu\nu}(x) = & U_\mu(x) U_\nu(x+\mu)U_\mu^\dagger(x+\nu) U_\nu^\dagger(x)\\\nonumber
& + U_\nu(x)U^\dagger_\mu(x-\mu+\nu)U^\dagger_\nu(x-\mu)U_\mu(x-\mu)\\\nonumber
& + U^\dagger_\mu(x-\mu)U^\dagger_\nu(x-\mu-\nu)U_\mu(x-\mu-\nu)U_\nu(x-\nu)\\
& + U^\dagger_\nu(x-\nu)U_\mu(x-\nu)U_\nu(x-\nu+\mu)U^\dagger_\mu(x).
\end{align}

Once operators have been constructed with the correct symmetry properties under $H(4)$, the lattice and continuum operators are related by a finite renormalisation factor
\begin{equation}
\mathcal{O}^{(E)}_{m,n} =  Z_2^m \mathcal{O}_{m,n}^{\text{latt.}},
\end{equation}
where the subscript $(m,n)$ denotes the $n$th vector from the $m$th representation, and $Z_2^m=1 + O(\alpha_s)$. The superscript $m$ on the renormalisation factor indicates that this can depend on representation. In this first investigation we do not compute $Z_2^m$, but note that for similar gluonic operators, such as the gluonic part of the energy momentum tensor, the corresponding renormalisation factor is $\mathcal{O}(1)$~\cite{Horsley:2012pz}. It would be surprising if $Z_2^m$, for any choice of $m$, was significantly different.

\renewcommand{\arraystretch}{1.5}
\begin{table*}
\centering
\begin{tabular}{cccc}\toprule
Rank & Operator & Symmetry & Dimension \\ \hline
$n+2$ & $G^{(E)}_{\mu\mu_1}\overleftrightarrow{D}^{(E)}_{\mu_3}\ldots \overleftrightarrow{D}^{(E)}_{\mu_n}G^{(E)}_{\nu \mu_2}$ & $\subset\underset{n+2}{\otimes} \tau_1^{(4)}$ & $n+2$ \\
$n$ & $G^{(E)}_{\mu_1\alpha} \overleftrightarrow{D}^{(E)}_{\mu_2}\ldots \overleftrightarrow{D}^{(E)}_{\mu_{n-1}}G^{(E)}_{\mu_n\alpha}$ & $\underset{n}{\otimes} \tau_1^{(4)}$ & $n+2$ \\
$n$ & $\epsilon_{\alpha\beta\gamma\mu_1}G^{(E)}_{\alpha\beta}\overleftrightarrow{D}^{(E)}_{\mu_2}\ldots \overleftrightarrow{D}^{(E)}_{\mu_{n-1}}G_{\mu_n\gamma}^{(E)}$ & $\tau_4^{(4)}\otimes\left(\underset{n-1}{\otimes} \tau_1^{(4)}\right)$ & $n+2$ \\
$n$ & $\overline{\psi}^{(E)} \gamma_{\mu_1}\gamma_5\overleftrightarrow{D}^{(E)}_{\mu_2}\ldots \overleftrightarrow{D}^{(E)}_{\mu_n}\psi^{(E)}$ & $\tau_4^{(4)}\otimes\left(\underset{n-1}{\otimes} \tau_1^{(4)}\right)$ & $n+2$ \\
$n$ & $\overline{\psi}^{(E)} \gamma_{\mu_1}\overleftrightarrow{D}^{(E)}_{\mu_2}\ldots \overleftrightarrow{D}^{(E)}_{\mu_n}\psi^{(E)}$ & $\underset{n}{\otimes} \tau_1^{(4)}$ & $n+2$ \\
$n$ & $\overline{\psi}^{(E)}\sigma_{\mu_1 \mu_2} \overleftrightarrow{D}^{(E)}_{\mu_3}\ldots\overleftrightarrow{D}^{(E)}_{\mu_n}\psi^{(E)}$ & $\subset\underset{n}{\otimes} \tau_1^{(4)}$ & $n+1$\\ \toprule
\end{tabular}
\caption{Dimensions and symmetry properties under $H(4)$ of operators that may mix with $\mathcal{O}^{(E)}_{\mu\nu{\mu_1}\ldots{\mu_{n}}}$. The symbol $\subset$ indicates that the operator transforms as a subset of the symmetry group shown.} \label{tab:1}
\end{table*}
\renewcommand{\arraystretch}{1}

\renewcommand{\arraystretch}{1.7}
\begin{table*}
\centering
\begin{tabular}{cll}\toprule
\hspace{2mm}Rank \hspace{2mm} & $\underset{m}{\otimes} \tau_1^{(4)}$& $\tau_4^{(4)}\otimes\left(\underset{m-1}{\otimes} \tau_1^{(4)}\right)$ \\ \hline
2 & $\tau _1^{(1)}$, $\tau _1^{(3)}$, $\tau _1^{(6)}$, $\tau _3^{(6)}$ & $\tau _4^{(1)}$, $\tau _4^{(3)}$, $\tau _1^{(6)}$, $\tau _4^{(6)}$\\
3 &  $\tau _2^{(4)}$, $\tau _4^{(4)}$,  $\tau _1^{(8)}$,  $\tau _2^{(8)}$&  $\tau _3^{(4)}$, $\tau _4^{(4)}$, $\tau _2^{(8)}$, $\tau _1^{(8)}$\\\
4 & $\tau _2^{(1)}$, $\tau _4^{(1)}$, $\tau _1^{(2)}$, $\tau _2^{(2)}$, $\tau _2^{(3)}$, \hphantom{666} & $\tau _3^{(1)}$, $\tau _1^{(1)}$, $\tau _2^{(2)}$, $\tau _1^{(2)}$, $\tau _3^{(3)}$ \\[-5pt] 
&  $\tau _3^{(3)}$, $\tau _4^{(3)}$, $\tau _2^{(6)}$, $\tau _4^{(6)}$ & $\tau _2^{(3)}$, $\tau _1^{(3)}$, $\tau _2^{(6)}$, $\tau _3^{(6)}$\\ \toprule
\end{tabular}
\caption{Irreducible representations which appear for the first time at each rank $m$ for the towers of operators in Table~\ref{tab:1}.}
\label{tab:irrepAppear}
\end{table*}
\renewcommand{\arraystretch}{1}

\subsection{Extraction of Results}
\label{sec:extract}

The expectation values of the matrix elements of the operators described in the previous section in the $\phi$ meson are extracted from ratios of two and three-point correlation functions. 
In order to compute these correlation functions, strange quark propagators were computed using a bare quark mass $m=-0.2450$ using 5 iterations of gauge-invariant Gaussian smearing~\cite{Gusken:1989qx} in the spatial directions at both source and sink. Measurements were performed for 96 different source locations on each of 1042 configurations, resulting in 100032 measurements. These propagators were contracted to form two-point and three-point $\phi$ meson correlators using interpolating operators of the form $\eta_i(x) =\overline{s}(x) \gamma_i s(x)$ in terms of smeared quark fields. 
For each type of correlator, measurements on each configuration were averaged and bootstrap statistical resampling was used in 
order to assess the statistical uncertainties in the measurements.
Note that the calculation does not include annihilation contributions (self-contraction of propagators at the source and sink), the effects of which are  OZI-suppressed.

The two point correlators
\begin{align}\nonumber
C^\text{2pt}_{jk}(t,\vec{p}) = & \sum_{\vec x} e^{i\vec{p}\cdot \vec{x}}\langle \eta_j(t,\vec{x})\eta^\dagger_k(0,\vec{0})\rangle \\ \nonumber
= & Z_\phi \left( e^{-Et} + e^{-E(T-t)}\right)\sum_{\lambda} \epsilon_j^{(E)}(\vec{p},\lambda)\epsilon^{(E)*}_k(\vec{p},\lambda)\\
& + \ldots, \label{eq:2pt}
\end{align}
were constructed  for all diagonal and off-diagonal polarisation combinations $(jk)$. 
The ellipsis denotes contributions from excited states.
For the spin-1 $\phi$ meson there are three different particle states such that
\begin{equation}\label{eq:EucPol}
\langle 0 | \eta_i (\vec{p}) | \vec{p}, \lambda \rangle = \sqrt{Z_\phi}\epsilon^{(E)}_i(\vec{p},\lambda)\,,
\end{equation}
where $\lambda=\{+,-,0\}$, and the polarisation vectors in Minkowski space have the explicit form
\begin{equation}
\epsilon^\mu(\vec{p},\lambda) = \left( \frac{\vec{p}\cdot\vec{e}_\lambda}{m}, \vec{e}_\lambda + \frac{\vec{p}\cdot\vec{e}_\lambda}{m(m+E)}\vec{p}\right),
\end{equation}
with $m$ and $E=\sqrt{|\vec{p}|^2+m^2}$ being the rest mass and energy of the state, and
\begin{align}
\vec{e}_\pm & = \mp \frac{1}{\sqrt{2}}(0,1,\pm i), \\
\vec{e}_0 & = (1,0,0).
\end{align}
The Euclidean polarisations needed for Eqs.~\eqref{eq:EucPol} and \eqref{eq:2pt} are
\begin{equation}
\epsilon_i^{(E)}(\vec{p},\lambda) = \epsilon^i(\vec{p},\lambda).
\end{equation}

To construct the three-point correlators corresponding to the insertion of the gluonic operator, the two point functions above were correlated configuration-by-configuration, and source-location--by--source-location, with the gluonic operator. 
The three-point correlators for a given operator $\mathcal{O}=\mathcal{O}_{m,n}^{\text{latt.}}$ have the form
\begin{align}\nonumber
C^\text{3pt}_{jk}(t,\tau,\vec{p})  = &  \sum_{\vec x}\sum_{\vec y} e^{i\vec{p}\cdot \vec{x}} \langle \eta_j(t,\vec{p}) \ \mathcal{O}(\tau,\vec{y})\ \eta_k^\dagger(0,\vec 0)\rangle \\  \nonumber
= & Z_\phi e^{-Et} \sum_{\lambda \lambda'} \epsilon_j^{(E)}(\vec{p},\lambda)\epsilon^{(E)*}_k(\vec{p},\lambda')\langle \vec{p},\lambda|\mathcal{O}|\vec{p},\lambda'\rangle \\ 
\label{eq:3pt}
& + \ldots 
\end{align}
if $0\ll\tau\ll t\ll T$ (where $T$ denotes the time extent of the lattice).  If we instead have $0\ll t\ll \tau\ll T$, $t$ is replaced by $(T-t)$ in the rightmost form of the above expression and there is an additional multiplicative factor of $(-1)^{n_4}$ where $n_4$ is the number of temporal indices in the operator $\mathcal{O}$.
In constructing $C^\text{3pt}$, various levels of  Wilson flow~\cite{Luscher:2010iy} or HYP smearing~\cite{Hasenfratz:2001hp} were applied to the links in the gluon operator. This was shown in Refs.~\cite{Meyer:2007tm,Alexandrou:2013tfa} to significantly improve the signal-to-noise ratio for a different gluon operator calculation.

Using Eq.~\eqref{eq:2pt} and Eq.~\eqref{eq:3pt}
we construct the ratio
\begin{equation}\label{eq:rat}
R_{jk}(t,\tau,\vec{p})=\frac{C^\text{3pt}_{jk}(t,\tau,\vec{p})+C^\text{3pt}_{jk}(T-t,T-\tau,\vec{p})}{C^\text{2pt}_{jk}(t,\vec{p})}
\end{equation}
for $\{t,\tau\}<T/2$. 
Other choices for the ratio, with different combinations of the two-point function in the denominator (e.g., spin-averaged) were also considered, and give consistent results.
This ratio may still depend on $t$ and $\tau$ due to contributions from higher states neglected in the derivation of Eq.~\eqref{eq:3pt}.
Note that the two point correlator in the denominator has reached its ground state after $t=8$.

To determine the dependence of the ratio in Eq.~\eqref{eq:rat} on the reduced matrix element $A_2$, we apply Eq.~\eqref{eq:expe} to the Minkowski-space versions of the Euclidean-space vectors in Appendix~\ref{sec:latticeExplicit}. The Minkowski operators are determined by 
noting that
\begin{align}
  G^{(E)}_{ij} & = G_{ij} \text{   if } i,j\in \{1,2,3\},\\
  G^{(E)}_{4j} & = (-i)G_{0j},
\end{align}
and so
\begin{equation}\label{eq:EtoM}
  \mathcal{O}_{m,n} \sim (-i)^{n_4}\mathcal{O}_{\mu\nu\mu_1 \mu_{2}},
\end{equation}
where $n_4$ is the number of temporal indices on the left-hand side, and temporal indices labelled `4' in Euclidean space correspond to indices labelled `0' in Minkowski space.

\begin{figure}
\begin{subfigure}[Cross-section at $\tau=4$.]{
\includegraphics[width=0.46\textwidth]{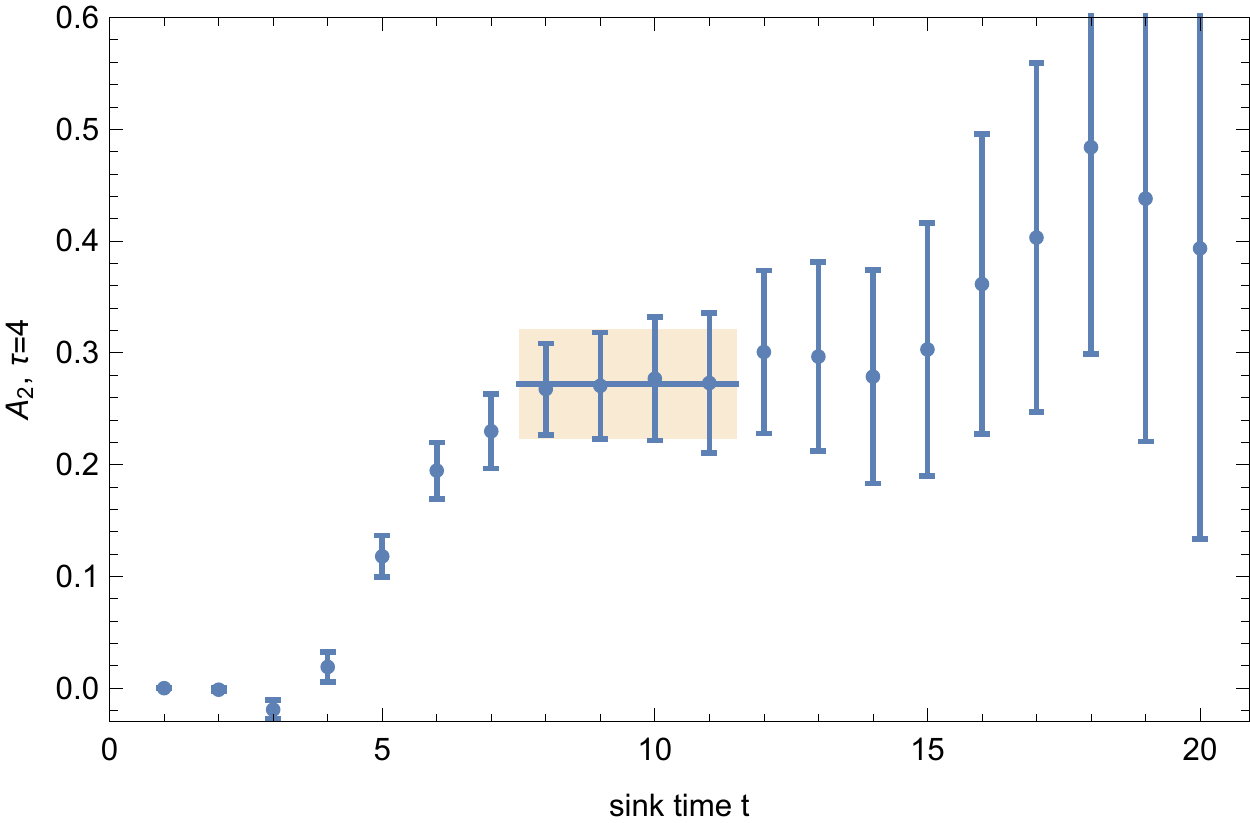}
}
\end{subfigure}
\begin{subfigure}[Cross-section at $t=9$.]{
\includegraphics[width=0.46\textwidth]{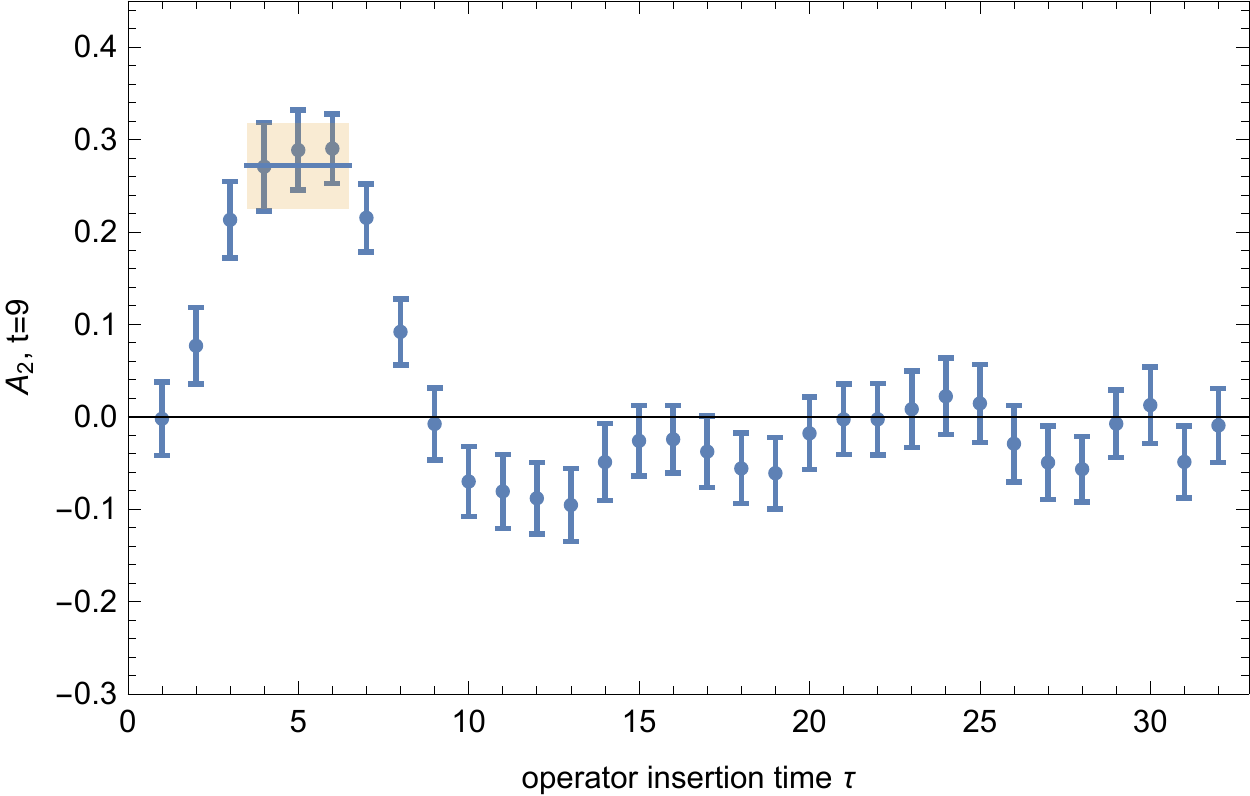}
}
\end{subfigure}
\caption{\label{fig:PlateauFig}Cross-sections of the plateau fit (in $t$ and $\tau$) to the reduced matrix element $A_2$ extracted from the ratio $R_{jk}(t,\tau,\vec{p})$ (Eq.~\eqref{eq:rat}) for the vector $\mathcal{O}^{(E)}_{1,1}$ at $|\vec{p}|^2=0$. Wilson flow~\cite{Luscher:2010iy} was applied to the links in the gluon operator as described in the text. 
}
\end{figure}

\begin{figure}
\includegraphics[width=0.44\textwidth]{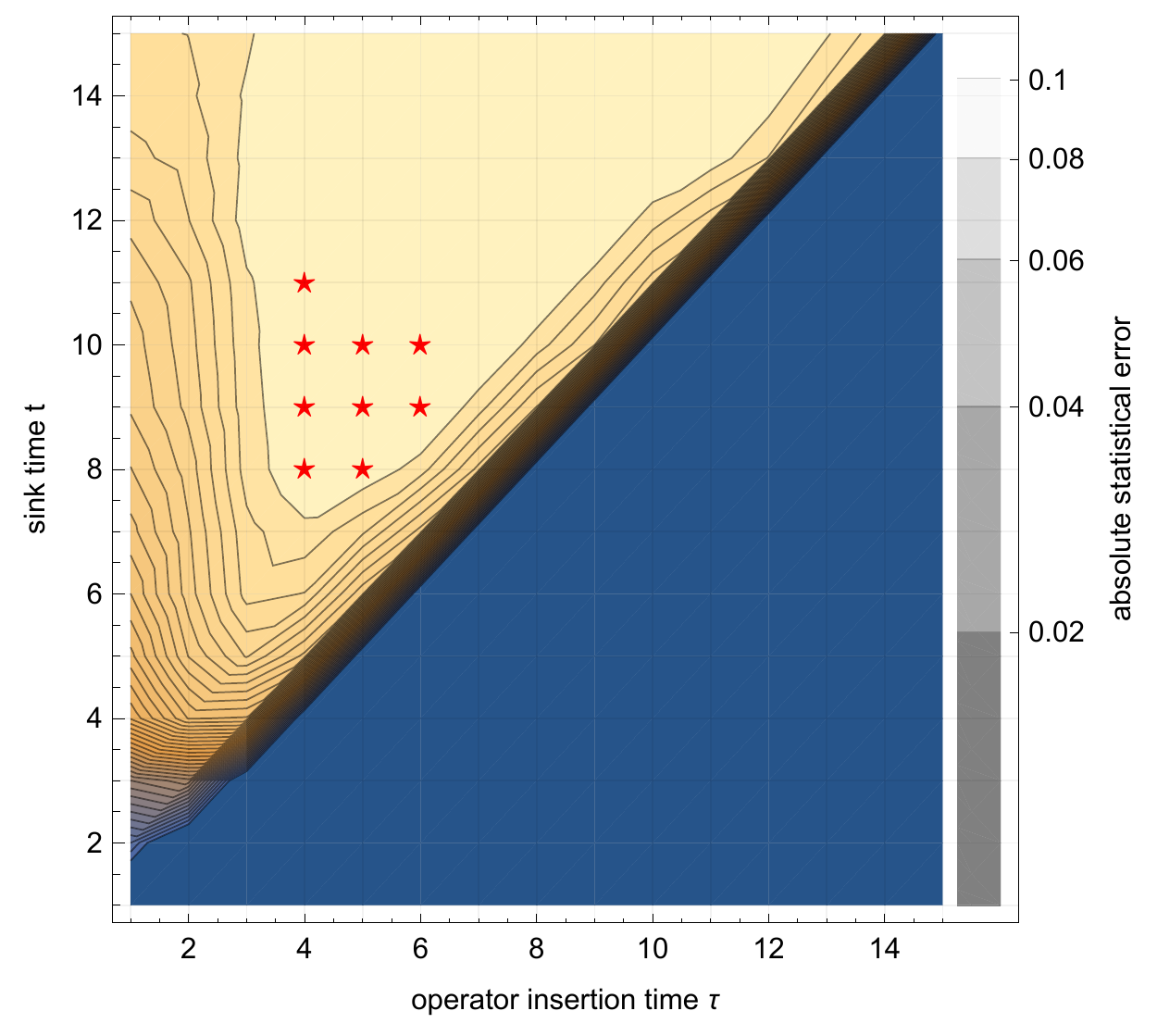}
\caption{\label{fig:ContourFig}Contour plot showing the fit region in the $t$--$\tau$ plane for the fits displayed in Fig.~\ref{fig:PlateauFig}. Each contour, moving out from the centre (i.e., moving from the pale to dark region), denotes an interval of one standard deviation from the central fit value. That is, results located on the third innermost contour are inconsistent with the fit result at 3 standard deviations. The results in the innermost pale region are consistent with the fit at 1 standard deviation. The red stars show the points included in the fit. Noise increases with increasing $t$ as illustrated in the vertical bar at the right of the figure which shows the $\tau$-averaged  ($0<\tau<t$) absolute statistical uncertainty of $A_2$.}
\end{figure}

\begin{figure}
\includegraphics[width=0.49\textwidth]{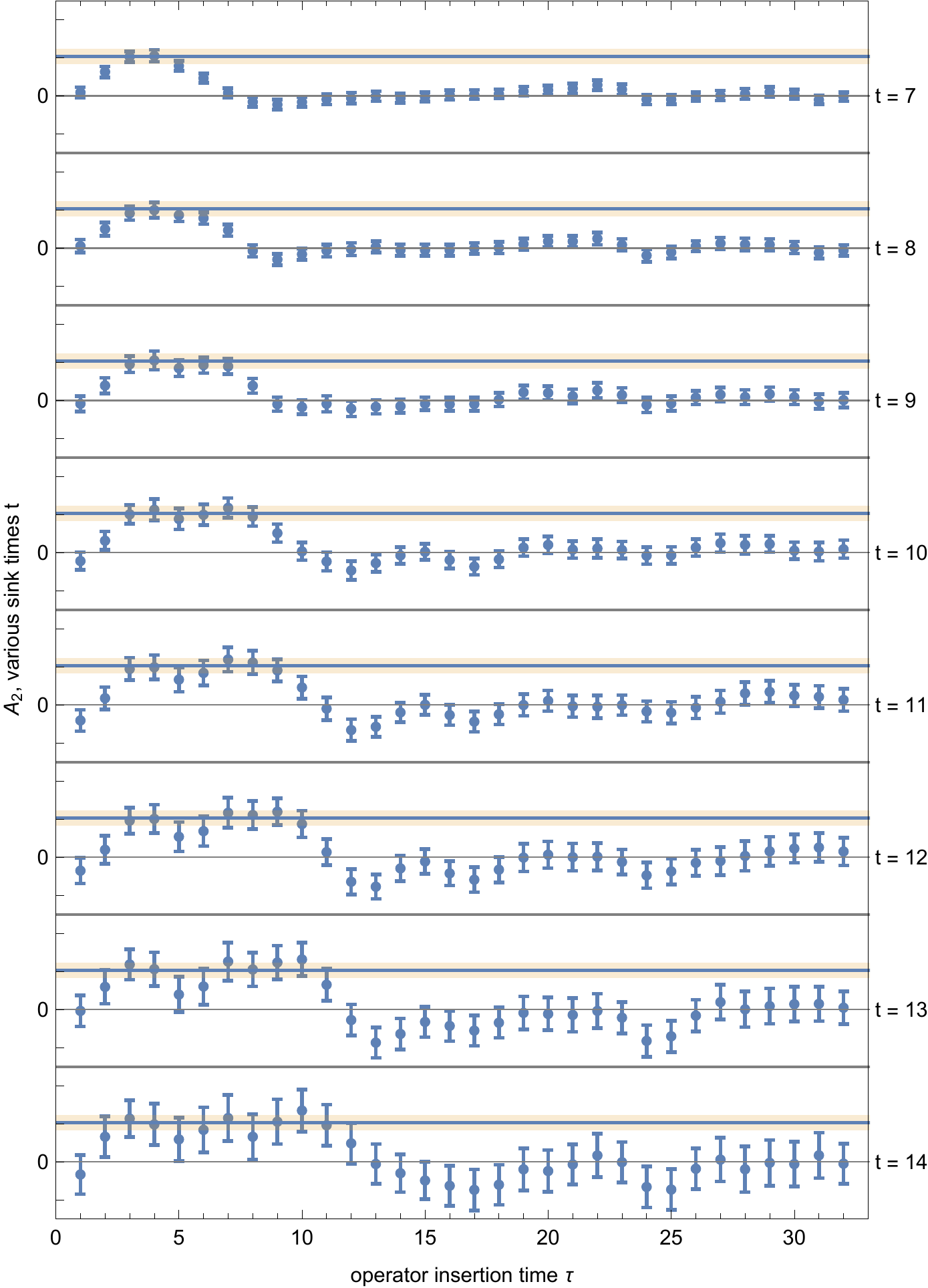}
\caption{\label{fig:BumpFig}Example of the evolution of the $\tau$-plateaus for $A_2$ with sink time $t$ for the vector $\mathcal{O}^{(E)}_{2,1}$ at $|\vec{p}|^2=3$. The horizontal bands show the final fit value obtained from the two dimensional ($t, \tau$) fit, as described in the text.}
\end{figure}

After averaging $R_{jk}(t,\tau,\vec{p})$, for a given basis vector $\mathcal{O}^{(E)}_{m,n}$, over the combinations of $\{j,k\}$ and equivalent boost momenta $\vec{p}$ which are non-zero by Eq.~\eqref{eq:expe}\footnote{Note that this averaging requires factors of the energy and three-momentum of the state; to determine the energy at a given $\vec{p}$, we use the measured mass and $E=\sqrt{|\vec{p}|^2+m^2}$.}, we determine plateaus in the $\tau$-dependence at each value of $t$, and in the $t$-dependence at each $\tau$, by searching for regions where the nearest-neighbour finite differences in $\tau$ or $t$ are consistent with zero. Taking the maximal connected overlap of the plateau regions defines a two-dimensional plateau in $\tau$ and $t$.
We perform a fit at the bootstrap level over that two-dimensional region to extract $A_2$. An example of the fit for the vector $\mathcal{O}^{(E)}_{1,1}$ (given explicitly in Appendix~\ref{sec:latticeExplicit}) at $|\vec{p}|^2=0$ is shown in Figs.~\ref{fig:PlateauFig} and \ref{fig:ContourFig}, and for the vector $\mathcal{O}^{(E)}_{2,1}$ at $|\vec{p}|^2=3$ is shown in Fig.~\ref{fig:BumpFig}.

\section{Soffer-like inequality}
\label{sec:soffer-like}

As well as the proof-of-principle extraction of $A_2$ described above, we also undertake a more general exploration of the gluonic structure of the $\phi$ meson. We consider here the gluonic analogue of the Soffer bound~\cite{Soffer:1994ww} for transversity in spin-1 particles. This bound, which is a positivity constraint, was first studied in moment space on the lattice in Ref.~\cite{Gockeler:2005cj,Diehl:2005ev}, for spin-1/2 particles.

The Soffer bound, for the isovector quark parton distribution functions in the proton, is
\begin{equation}
|\delta q(x)|\le \frac{1}{2}\left(q(x)+\Delta q(x)\right),
\end{equation}
where $q(x)$, $\Delta q(x)$ and $\delta (x)$ denote the spin-independent, spin-dependent and transversity quark distributions and we suppress the renormalisation-scale dependence. The gluonic analogue of this expression, for spin-1 particles, is~\cite{Artru:2008cp,Bob,Boer:2016xqrx,Cotogno:2017mwy} 
\begin{equation}\label{eq:sofferdist}
|\delta G(x)|\le \frac{1}{2}\left(f_1(x)+\frac{1}{2}f_{1LL}(x) + g_1(x)\right),
\end{equation}
where $\delta G (x)$ is the gluonic transversity distribution defined in Eq.~\eqref{eq:deltaG}, $f_1(x)$ and $f_{1LL}(x)$ are the spin-independent gluon distributions, and $g_1(x)$ is the spin-dependent gluon distribution. The notation here for $f_1$, $f_{1LL}$ and $g_1$ is the same as in Refs.~\cite{Boer:2016xqrx,Cotogno:2017mwy}, while $\delta G(x)$ is named $h_{1TT}$ in those works.
The Mellin moments of this equation are related to the reduced matrix elements of local operators:
\begin{align}
\mathcal{O}_{\mu\nu\mu_1 \ldots \mu_{n}} = & S\left[G_{\mu\mu_1} \overleftrightarrow{D}_{\mu_3}\ldots  \overleftrightarrow{D}_{\mu_{n}}G_{\nu \mu_2}\right], \\ \label{eq:SDgluonop}
\overline{\mathcal{O}}_{\mu_1 \ldots \mu_{n}} = & S\left[G_{\mu_1\alpha} \overleftrightarrow{D}_{\mu_3}\ldots \overleftrightarrow{D}_{\mu_{n}}G_{\mu_2}^{\ \ \alpha}\right], \\
\widetilde{\mathcal{O}}_{\mu_1 \ldots \mu_{n}} = & S\left[\widetilde{G}_{\mu_1\alpha} \overleftrightarrow{D}_{\mu_3}\ldots \overleftrightarrow{D}_{\mu_{n}}G_{\mu_2}^{\ \ \alpha}\right],
\end{align}
for the transversity, spin-independent and spin-dependent distributions respectively, where the dual field strength tensor is $\widetilde G_{\mu\nu}=\frac{1}{2}\epsilon_{\mu\nu\rho\sigma} G^{\rho\sigma}$. The first moments of the gluonic distributions are related to the matrix elements of the $n=2$ operators in the towers above.
Since the $\phi$ matrix element of $\widetilde{\mathcal{O}}_{\mu_1\mu_2}$ vanishes by operator symmetries, the analogue of the Soffer bound for the leading Mellin moments of gluon distributions is~\cite{Cotogno:2017mwy}
\begin{equation}\label{eq:Soffer}
|A_2|\le \frac{1}{24} (5 B_{2,1} - 6 B_{2,2}),
\end{equation}
where $A_2$ is the reduced matrix element defined in Eq.~\eqref{eq:expe} and $B_{2,1}$ and $B_{2,2}$ are linear combinations of the moments of the structure functions $f_1$ and $f_{1LL}$ in Eq.~\eqref{eq:sofferdist}, defined through
\begin{align}\nonumber
\langle p E' | \overline{\mathcal{O}}_{\mu_1\mu_2}&|p E\rangle \\\nonumber
=&S\left[M^2E'^{*}_{\mu_1}E_{\mu_2}\right]B_{2,1}(\mu^2)\\\label{eq:ObarME}
&+ S \left[(E\cdot E'^{*})p_{\mu_1}p_{\mu_2} \right]B_{2,2}(\mu^2).
\end{align}

The building block of the Euclidean analogue of Eq.~\eqref{eq:SDgluonop} for $n=2$ is
\begin{equation}
\overline{\mathcal{O}}_{\mu_1\mu_2}=G_{\mu_1\alpha}^{(E)}G_{\mu_2\alpha}^{(E)}.
\end{equation}
It is clear from Table~\ref{tab:1} that this operator is subject to mixing with same-dimension quark operators at $\mathcal{O}(\alpha_s)$. In this proof-of-principle study we neglect operator mixing and renormalisation and simply determine the the bare coefficients 
in Eq.~\eqref{eq:ObarME}, $B_{2,1}$ and $B_{2,2}$, as described in previous sections for $A_2$, from the matrix elements of Euclidean-space basis vectors in appropriate irreducible representations of $H(4)$. Explicit forms for the particular vectors we consider are given in Appendix~\ref{sec:latticeExplicit}.

\section{Results}

\begin{figure}
\includegraphics[width=0.49\textwidth]{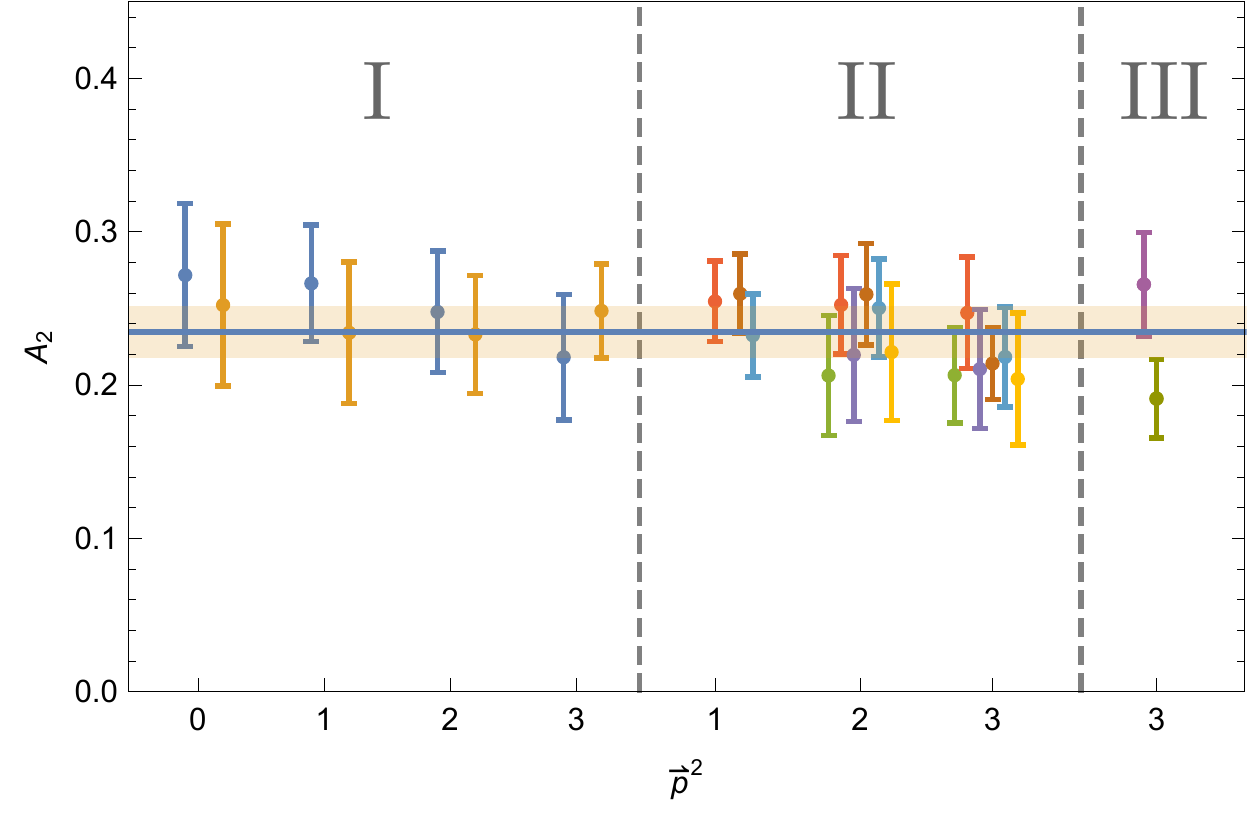}
\caption{\label{fig:ResultFig}Reduced matrix element $A_2$ extracted from ratios of two and three-point functions for different boost momenta, as described in Section~\ref{sec:extract}. Wilson flow~\cite{Luscher:2010iy} was applied to the links in the gluon operator as described in the text. Results in sections I, II and III of the figure are determined from vectors in the $\tau_1^{(2)}$, $\tau_1^{(6)}$ and  $\tau_2^{(2)}$ representations. Different colours (offset on the horizontal access for clarity) denote different vectors in each basis. The horizontal band is a fit shown to guide the eye.}
\end{figure}

The reduced matrix element $A_2$ obtained from this analysis, with Wilson flow~\cite{Luscher:2010iy} applied to the links in the gluon operator to a total flow time of 1 in lattice units using a step size of 0.01, is shown in Fig.~\ref{fig:ResultFig} for various boosts and for all operator basis vectors that have non-vanishing contributions at that boost. Outstanding agreement is seen between the values obtained using vectors in the three different irreducible representations considered, as well as between measurements of different basis vectors within each irreducible representation. The values in different irreducible representations are expected to differ by both lattice discretisation artefacts and because of differences in the renormalisation constants. Their agreement suggests that such effects are not severe at the lattice spacing used in this study. There is also no difference, within uncertainties, between results with Wilson-flowed gauge fields and results that instead use 2 or 5 steps of HYP (hypercubic \cite{Hasenfratz:2001hp}) smearing (shown in Appendix~\ref{sec:otherResults}), even though each set of results will have a different renormalisation factor.

It is significant that we find a statistically clean and theoretically consistent signal in this unphysical simulation. Nevertheless, from these unrenormalised results, at a single lattice spacing, on a single lattice volume and with a single choice of quark masses, we can only draw fairly imprecise conclusions about the size of $A_2$. 
Performing a constant fit to all extractions for the Wilson-flowed data, and assigning a conservative 20\% uncertainty due to the missing renormalisation (which is in fact significantly larger than the difference between results obtained using different levels of smearing or Wilson flow), we find $A_2 \sim0.23(2)(5)$, where the first uncertainty is statistical and the second is an estimate of renormalisation effects.
This result bodes extremely well for the application of the methods described here to future improved studies with multiple lattice spacings and physical quark masses as well as for the more phenomenologically relevant calculations of $A_2$ in light nuclei and of the off-forward nucleon matrix elements of the gluonic transversity operator.

\begin{figure}
\includegraphics[width=0.49\textwidth]{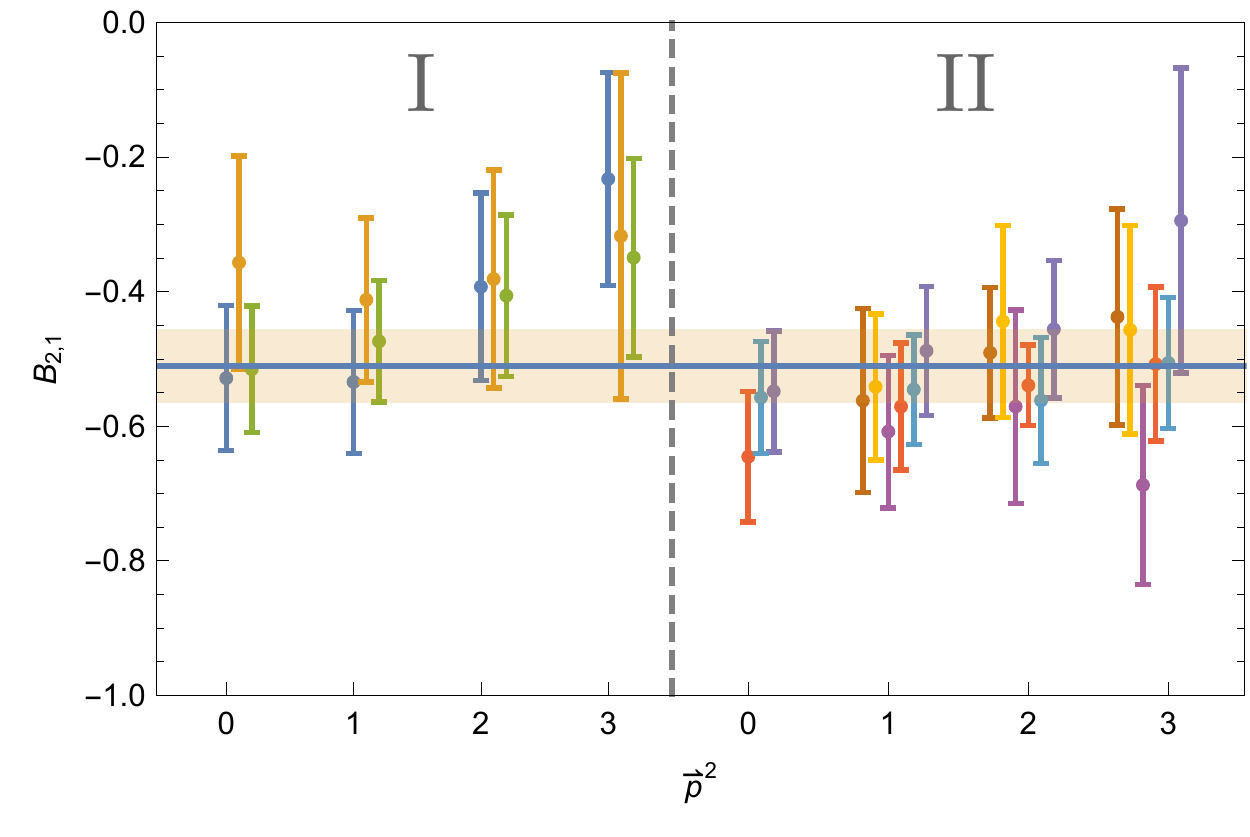}
\includegraphics[width=0.49\textwidth]{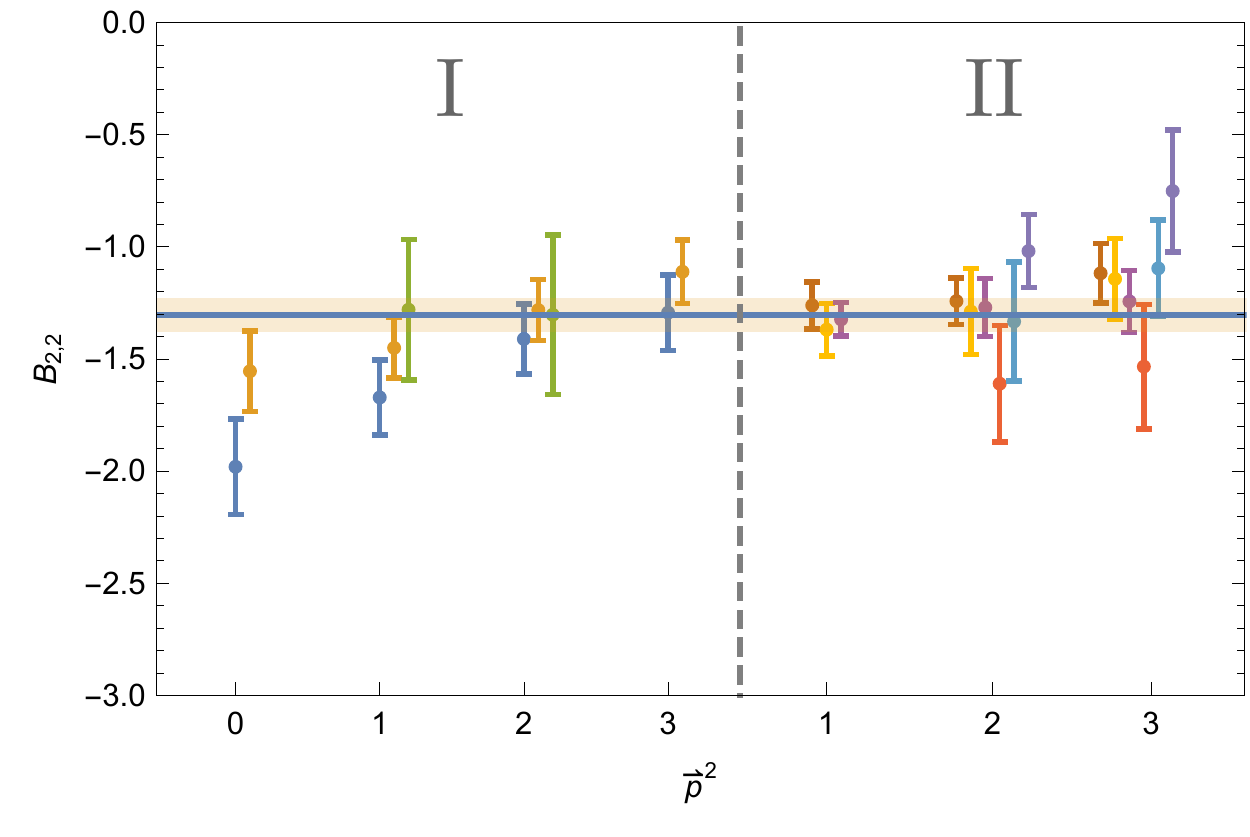}
\caption{\label{fig:ResultFigF2}Reduced matrix elements $B_{2,1}$ and $B_{2,2}$ extracted from ratios of two and three point functions as described in Sections~\ref{sec:extract} and \ref{sec:soffer-like}. Wilson flow~\cite{Luscher:2010iy} was applied to the links in the gluon operator as described in the text. Results in sections I and II of the figure are determined using vectors in the $\tau_1^{(3)}$ and $\tau_3^{(6)}$ representations. Different colours (offset on the horizontal access for clarity) denote different basis vectors. The horizontal band is a fit shown to guide the eye.}
\end{figure}

Results for the reduced matrix element $B_2$, associated with the spin-independent gluon distribution, are shown in Figure~\ref{fig:ResultFigF2}. Clearly we once again find excellent agreement between results determined using different Euclidean basis vectors and different irreducible representations, as well as between the different levels of HYP smearing and Wilson flow that we consider (results for 2 and 5 steps of HYP smearing are shown in Appendix~\ref{sec:otherResults}).
Renormalised and unmixed operators would be needed to make a concrete statement about the Soffer bound (Eq.~\eqref{eq:Soffer}) in this context. However, one might expect the renormalisation factors and a number of other statistical and systematic uncertainties to cancel to some extent in the expression in Eq.~\eqref{eq:Soffer}.
Under the assumptions of small mixing effects in $B_{2,1}$ and $B_{2,2}$ and approximately equal renormalisation constants for $A_2$, $B_{2,1}$ and $B_{2,2}$, the results shown in Figs.~\ref{fig:ResultFig} and \ref{fig:ResultFigF2} suggest that the Soffer bound for the first moment of the gluon distributions is saturated to about 80\%--100\%, similar to the results found in a lattice study of the bound for the first two moments of the quark distributions of the nucleon~\cite{Gockeler:2005cj,Diehl:2005ev}.

\section{Conclusion}

Despite the ubiquity of gluons in QCD, understanding the gluonic structure of hadrons and nuclei is considerably more difficult than understanding the quark structure. In part, this can be attributed to the significant experimental challenges inherent in measurements of gluon observables that are typically ${\cal O}(\alpha_s)$-suppressed relative to quark observables. The proposed Electron-Ion Collider~\cite{Accardi:2012qut} is designed to focus on these challenges and measure a wide range of gluon observables in hadrons and in nuclei. In this work, we have performed proof-of-principle calculations demonstrating the utility of lattice QCD in providing benchmarks for such experiments. In particular, we have  studied the first moment of the leading-twist, double-helicity-flipping gluon transversity structure function $\Delta (x,Q^2)$ in the $\phi$ meson by computing the matrix element of the local twist-2 operator in Eq.~(\ref{eq:OE}). It is significant that we find a statistically clean and theoretically consistent and robust signal in this unphysical simulation. For example, statistical agreement is seen between the values of this observable obtained using vectors in three different irreducible representations to which the operator under consideration subduces at nonzero lattice spacing, as well as between measurements of different basis vectors within each irreducible representation. These values are expected to differ by lattice discretisation artefacts, suggesting that such effects are not severe. There is also agreement between the results with different levels of HYP smearing and with Wilson-flowed gauge fields, even though each set of results has a different renormalisation. 
In addition, we have explored the gluonic analogue of the Soffer bound for transversity for the first time, showing that the first moment of this bound in a $\phi$ meson (at the unphysical light quark masses used in this work and subject to caveats regarding renormalisation and the continuum limit) is saturated to approximately the same extent as the first moment of the isovector quark Soffer bound for the nucleon as determined in a previous lattice simulation~\cite{Gockeler:2005cj,Diehl:2005ev}. 

 This study is encouraging for the application of the methods described here to the calculation of off-forward gluonic transversity matrix elements in the nucleon. These quantities determine moments of gluon generalised parton distributions that are accessible in DVCS. It is also encouraging for calculations of moments of $\Delta (x,Q^2)$ in light nuclei, where this structure function provides a measure of exotic glue---the contributions from gluons not associated with individual nucleons in a nucleus. 
 While nuclei are considerably more challenging to study in lattice QCD than simple hadrons like the $\phi$ meson, there has been considerable recent progress on lattice studies of the spectroscopy~\cite{Yamazaki:2009ua,Beane:2011iw,Beane:2012vq} and properties~\cite{Beane:2014ora, Chang:2015qxa} of light nuclei. Although a procedure to measure $\Delta (x,Q^2)$ in nuclei was first outlined in 1989~\cite{Jaffe:1989xy}, it is only recently in a letter of intent to Jefferson Lab~\cite{JLAB.LOI} that an experimental measurement of $\Delta(x,Q^2)$ has been proposed, with the goal of measurements at low $x$ on nitrogen targets. Further measurements could be expected at a future Electron-Ion Collider~\cite{Accardi:2012qut,Kalantarians:2014eda}.

\section*{Acknowledgements}
This work is supported by the U.S. Department of Energy under Early Career Research Award DE-SC0010495 and under Grant No. DE-SC0011090. The Chroma software library~\cite{Edwards:2004sx} was used in the data analysis. PES thanks the Institute for Nuclear Theory at the University of Washington for its hospitality. We thank Ian Clo\"et, Bob Jaffe, James Maxwell and Richard Milner for helpful discussions.
We thank Kostas Orginos for production of the gauge configurations used in this work.

\appendix

\section{Explicit Lattice Basis Vectors}
\label{sec:latticeExplicit}

Here we list the explicit forms of the Euclidean basis vectors which were used for the calculations described in Section~\ref{sec:lattice}.

The Euclidean analogue of the operator defined in Eq.~\eqref{eq:O2} is built from
\begin{equation}
\mathcal{O}^{(E)}_{\mu\nu\mu_1 \mu_{2}} = G^{(E)}_{\mu\mu_1}G^{(E)}_{\nu \mu_2}.
\end{equation}
We consider the $\tau_1^{(2)}$, $\tau_2^{(6)}$ and $\tau_2^{(2)}$ irreducible representations.
For $\tau_1^{(2)}$, the basis vectors are~\cite{Gockeler:1996mu}:
\begin{align}\nonumber
\mathcal{O}_{{1,1}}^{(E)} = & \frac{1}{8\sqrt{3}}\left(-2 {\mathcal{O}}^{(E)}_{1122}+{\mathcal{O}}^{(E)}_{1133}+{\mathcal{O}}^{(E)}_{1144} \right.\\
 & \left. \hphantom{\frac{1}{8\sqrt{3}}\,\,\,\,\,} +{\mathcal{O}}^{(E)}_{2233}+{\mathcal{O}}^{(E)}_{2244}-2 {\mathcal{O}}^{(E)}_{3344} \right), \\
\mathcal{O}_{{1,2}}^{(E)} = & \frac{1}{8} \left({\mathcal{O}}^{(E)}_{1144}+{\mathcal{O}}^{(E)}_{2233}-{\mathcal{O}}^{(E)}_{1133}-{\mathcal{O}}^{(E)}_{2244}\right).
\end{align}

The $\tau_2^{(6)}$ vectors are:
\begin{align}
\mathcal{O}_{{2,1}}^{(E)} = & \frac{1}{4} \left({\mathcal{O}}^{(E)}_{1123}-{\mathcal{O}}^{(E)}_{2344}\right),\\
\mathcal{O}_{{2,2}}^{(E)} = & \frac{1}{4} \left({\mathcal{O}}^{(E)}_{1124}+{\mathcal{O}}^{(E)}_{2334}\right),\\
\mathcal{O}_{{2,3}}^{(E)} = & \frac{1}{4} \left({\mathcal{O}}^{(E)}_{1223}+{\mathcal{O}}^{(E)}_{1344}\right),\\
\mathcal{O}_{{2,4}}^{(E)} = & \frac{1}{4} \left({\mathcal{O}}^{(E)}_{1224}-{\mathcal{O}}^{(E)}_{1334}\right),\\
\mathcal{O}_{{2,5}}^{(E)} = & \frac{1}{4} \left({\mathcal{O}}^{(E)}_{1134}-{\mathcal{O}}^{(E)}_{2234}\right),\\
\mathcal{O}_{{2,6}}^{(E)} = & \frac{1}{4} \left({\mathcal{O}}^{(E)}_{1233}-{\mathcal{O}}^{(E)}_{1244}\right).
\end{align}

Finally, we consider the $\tau_2^{(2)}$ basis vectors:
\begin{align}
\mathcal{O}_{{3,1}}^{(E)} = & \frac{1}{4} \left({\mathcal{O}}^{(E)}_{1324}+{\mathcal{O}}^{(E)}_{1234}\right),\\
\mathcal{O}_{{3,2}}^{(E)} = & 4\sqrt{3} \left({\mathcal{O}}^{(E)}_{1324}-{\mathcal{O}}^{(E)}_{1234}-2{\mathcal{O}}^{(E)}_{1243}\right).
\end{align}

To construct the Euclidean analogue of Eq.~\eqref{eq:SDgluonop}, we use
\begin{equation}
\overline{\mathcal{O}}^{(E)}_{\mu_1\mu_2}=G_{\mu_1\alpha}^{(E)}G_{\mu_2\alpha}^{(E)}.
\end{equation}
Two irreducible representations are considered here. 
For the $\tau_1^{(3)}$ representation the basis vectors are:
\begin{align}
\overline{\mathcal{O}}_{{1,1}}^{(E)} = & \frac{1}{2}\left(\overline{\mathcal{O}}^{(E)}_{11}+\overline{\mathcal{O}}^{(E)}_{22}-\overline{\mathcal{O}}^{(E)}_{33}-\overline{\mathcal{O}}^{(E)}_{44}\right), \\
\overline{\mathcal{O}}_{{1,2}}^{(E)} = & \frac{1}{\sqrt{2}} \left(\overline{\mathcal{O}}^{(E)}_{33}-\overline{\mathcal{O}}^{(E)}_{44}\right),\\
\overline{\mathcal{O}}_{{1,3}}^{(E)} = & \frac{1}{\sqrt{2}} \left(\overline{\mathcal{O}}^{(E)}_{11}-\overline{\mathcal{O}}^{(E)}_{22}\right).
\end{align}

For $\tau_3^{(6)}$ the vectors are:
\begin{align}
\overline{\mathcal{O}}_{{2,\mu\nu}}^{(E)} = & \frac{1}{\sqrt{2}} \left(\overline{\mathcal{O}}^{(E)}_{\mu\nu}+\overline{\mathcal{O}}^{(E)}_{\nu\mu}\right), \hspace{2mm} 1\le \mu < \nu \le 4.
\end{align}

\section{Results with HYP smearing}
\label{sec:otherResults}

Here we show the results of our analysis with 2 and 5 steps of HYP smearing (rather than with Wilson flow as in the main text). Clearly, for each choice of smearing, there is agreement between the values obtained using vectors in the different irreducible representations considered, as well as between measurements of different basis vectors within each irreducible representation.

\begin{figure*}[p]
\begin{subfigure}[ Two steps of HYP smearing]{
\includegraphics[width=0.48\textwidth]{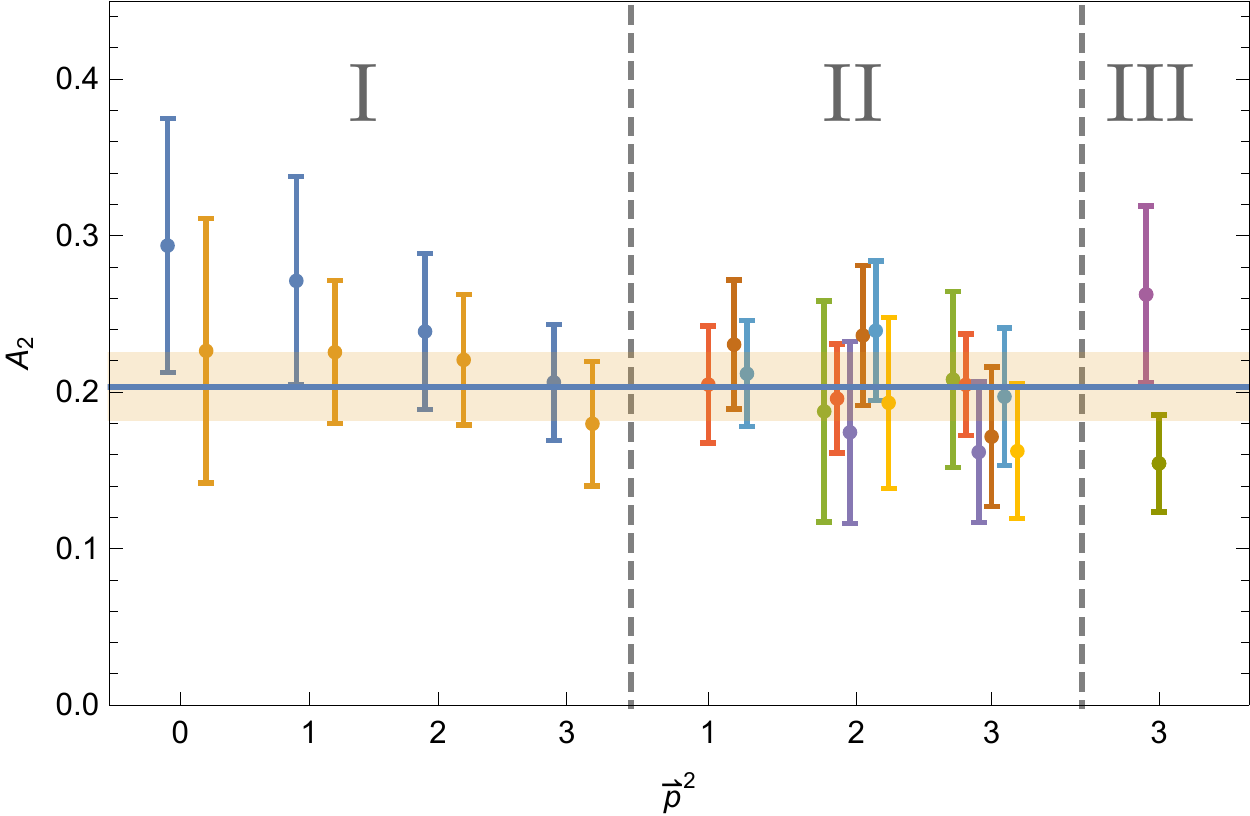}}
\end{subfigure}
\begin{subfigure}[ Five steps of HYP smearing]{
\includegraphics[width=0.48\textwidth]{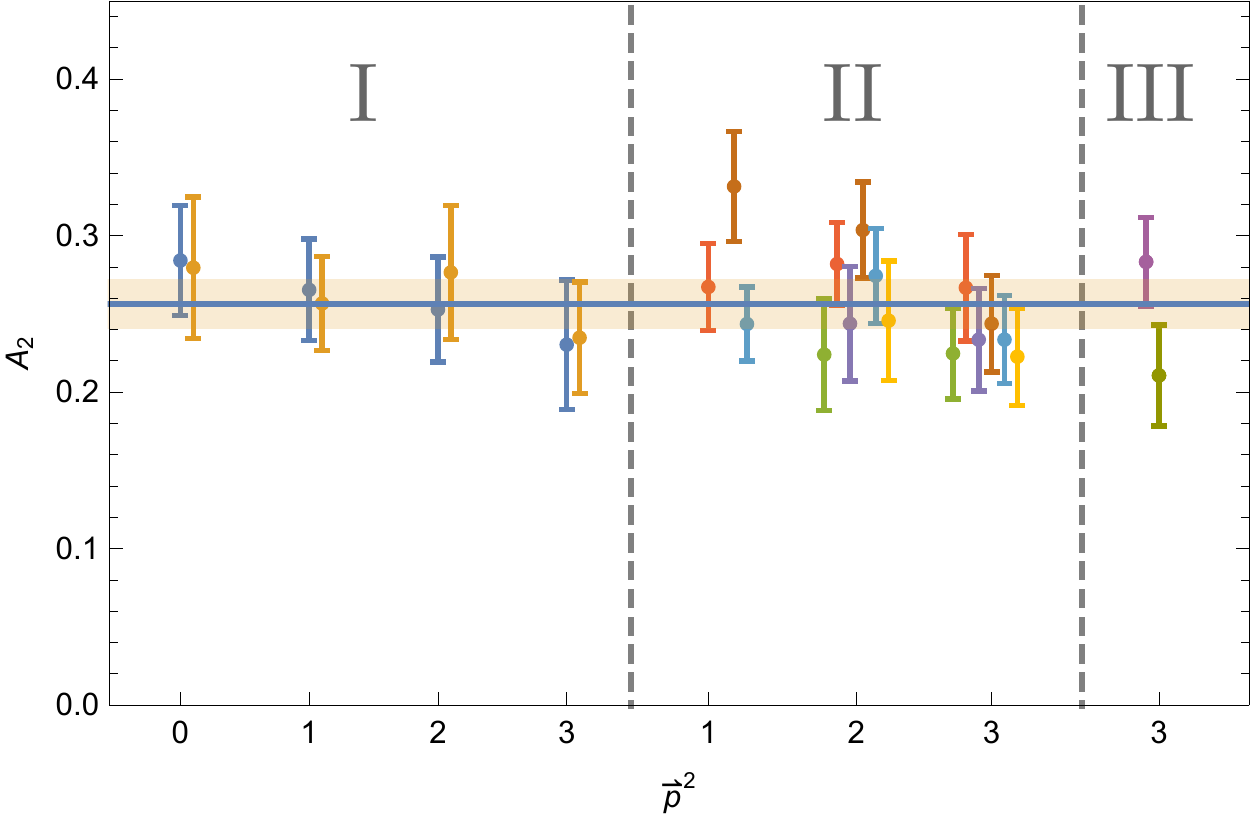}}
\end{subfigure}
\caption{\label{fig:ResultFig2}Reduced matrix element $A_2$ extracted from ratios of two and three-point functions as described in Section~\ref{sec:extract}. Results in sections I, II and III of the figure are determined from vectors in the $\tau_1^{(2)}$, $\tau_1^{(6)}$ and  $\tau_2^{(2)}$ representations. Different colours (offset on the horizontal access for clarity) denote different vectors in each basis. These results can be compared with those in Fig.~\ref{fig:ResultFig} with Wilson-flowed gauge fields. The horizontal bands are fits shown to guide the eye.}
\end{figure*}

\begin{figure*}[p]
\begin{subfigure}[ Two steps of HYP smearing]{
\includegraphics[width=0.48\textwidth]{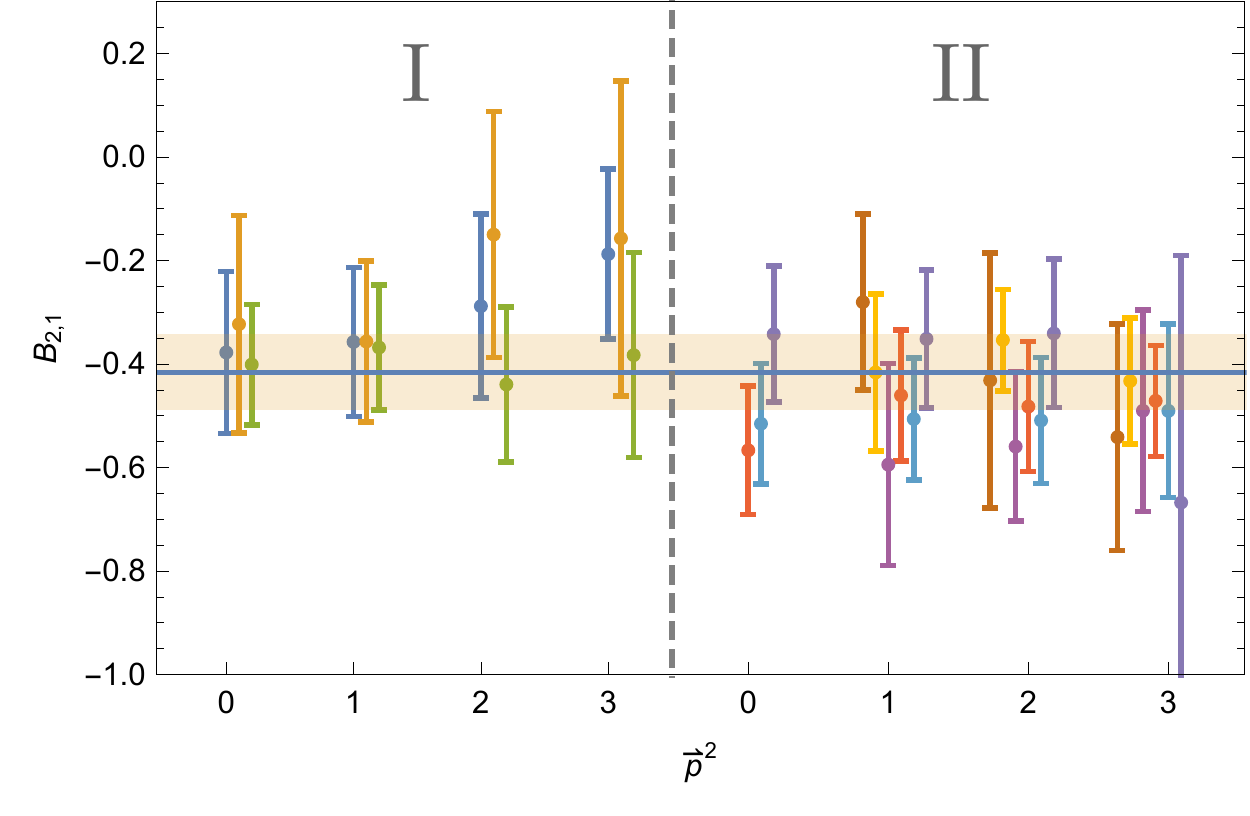}}
\end{subfigure}
\begin{subfigure}[ Five steps of HYP smearing]{
\includegraphics[width=0.48\textwidth]{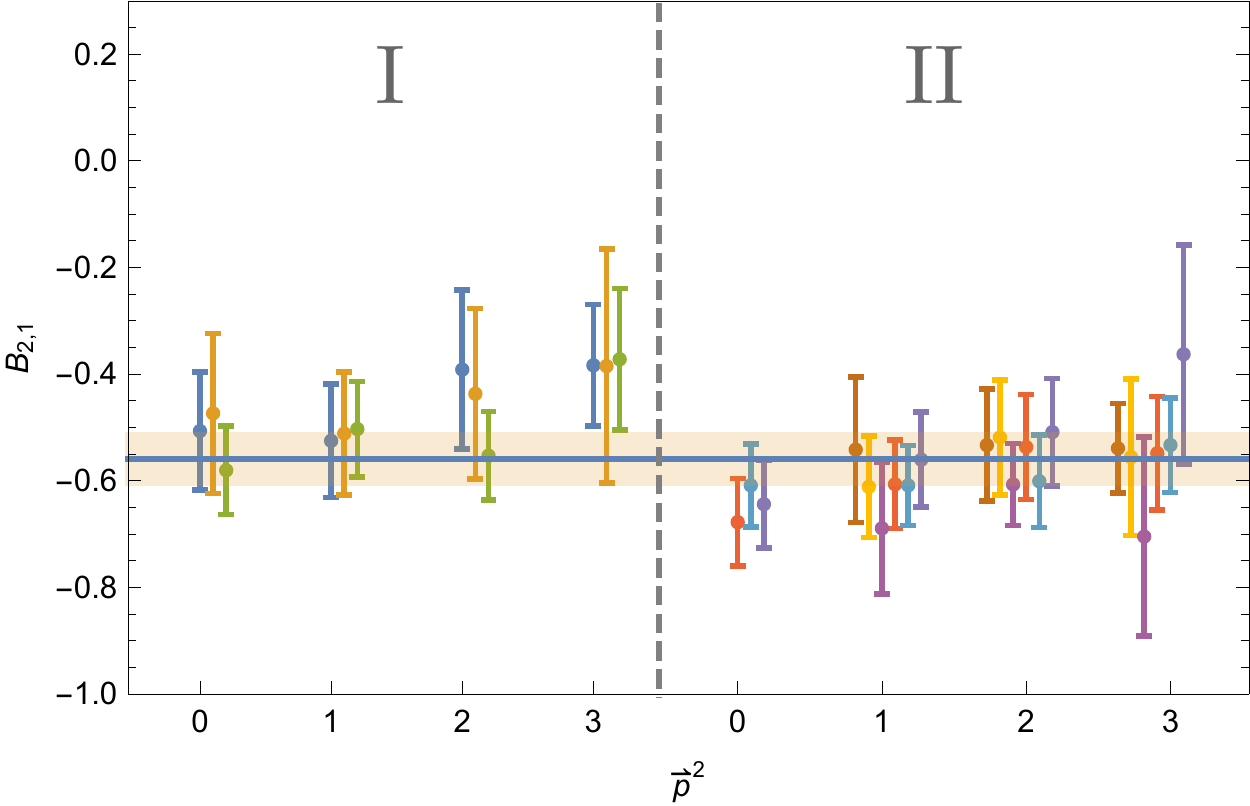}}
\end{subfigure}
\begin{subfigure}[ Two steps of HYP smearing]{
		\includegraphics[width=0.48\textwidth]{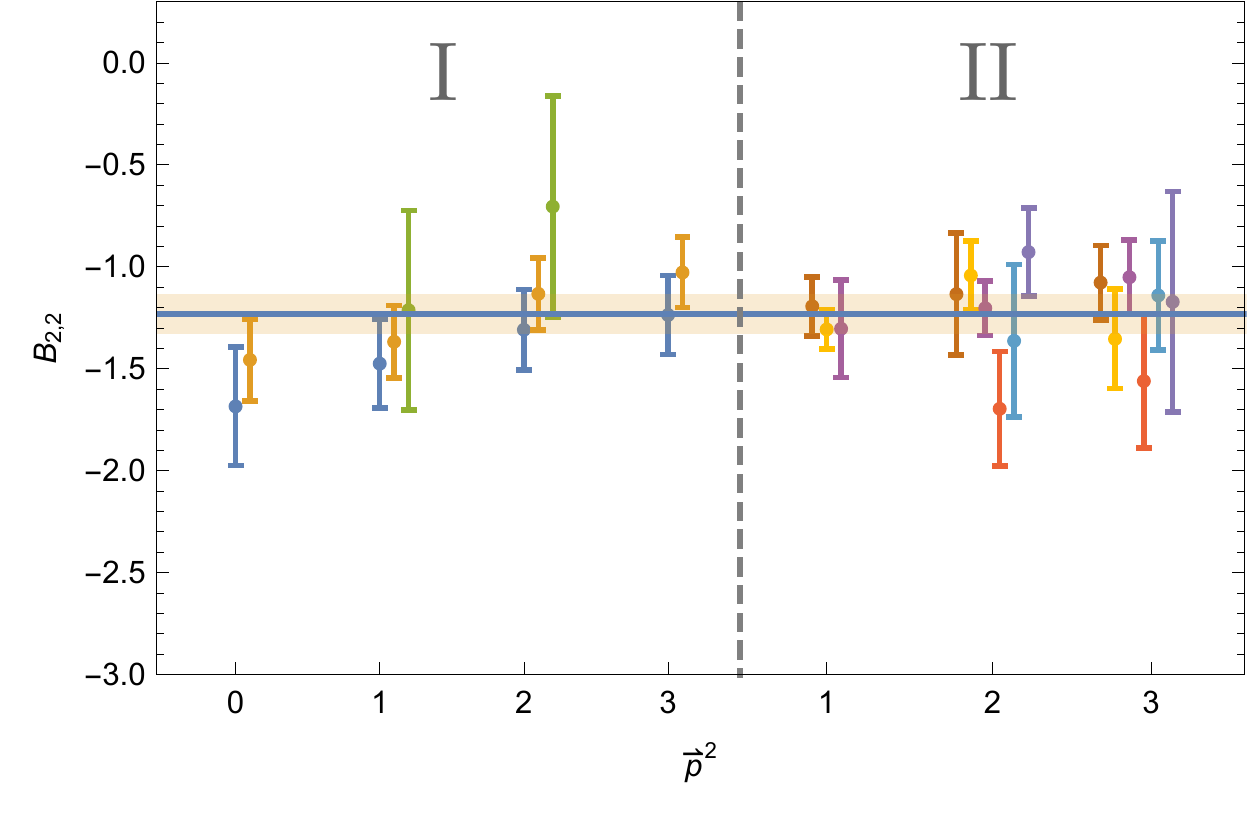}}
\end{subfigure}
\begin{subfigure}[ Five steps of HYP smearing]{
		\includegraphics[width=0.48\textwidth]{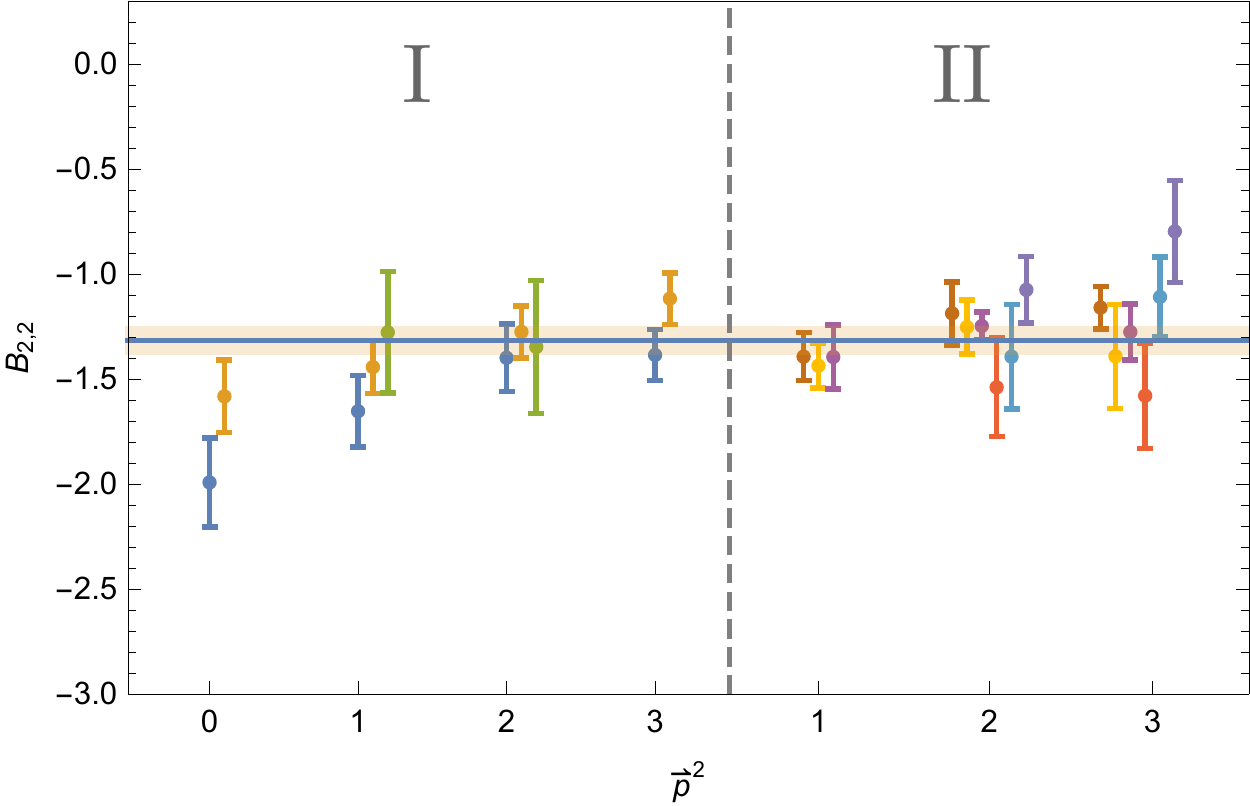}}
\end{subfigure}
\caption{\label{fig:ResultFigF22}Reduced matrix elements $B_{2,1}$ and $B_{2,2}$ extracted from ratios of two and three-point functions as described in Sections~\ref{sec:extract} and \ref{sec:soffer-like}. Results in sections I and II are determined using vectors in the $\tau_1^{(3)}$ and $\tau_3^{(6)}$ representations. Different colours (offset on the horizontal access for clarity) denote different basis vectors. These results can be compared with those in Fig.~\ref{fig:ResultFigF2} with Wilson-flowed gauge fields. The horizontal bands are fits shown to guide the eye.}
\end{figure*}

\bibliography{DGBib}

\end{document}